\documentclass[pdflatex,sn-basic]{sn-jnl}% Math and Physical Sciences Numbered Reference Style

\usepackage{graphicx}%
\usepackage{multirow}%
\usepackage{amsmath,amssymb,amsfonts}%
\usepackage{amsthm}%
\usepackage{mathrsfs}%
\usepackage[title]{appendix}%
\usepackage{xcolor}%
\usepackage{textcomp}%
\usepackage{manyfoot}%
\usepackage{booktabs}%
\usepackage{algorithm}%
\usepackage{algorithmicx}%
\usepackage{algpseudocode}%
\usepackage{listings}%
\usepackage{url}
\usepackage{tikz}
\usetikzlibrary{patterns}

\theoremstyle{thmstyleone}%

\theoremstyle{thmstyletwo}%

\theoremstyle{thmstylethree}%
\newtheorem{definition}{Definition}%

\begin{document}

%% --- Custom commands ---------------------------------------------------
\newcommand{\dif}{\mathrm{d}}        % differential; do NOT use \dd (reserved)
\newcommand{\QQ}{\mathbb{Q}}         % risk-neutral measure
\newcommand{\PP}{\mathbb{P}}         % physical measure
\newcommand{\Ebb}{\mathbb{E}}        % expectation
\newcommand{\Rbb}{\mathbb{R}}        % real numbers
\newcommand{\phinet}{\phi_t}         % basis factor
\newcommand{\TAB}{\widehat{\phi}}    % filtered basis

%% --- Theorem env not predefined by sn-jnl ------------------------------
\newtheorem{assumption}{Assumption}

\setlength{\parindent}{1.5em}
\setlength{\parskip}{0.15em}
\linespread{1.05}

% === FRONT MATTER ===================================================
\title[Token-to-Fiat Admissibility]{Price-Discovery Admissibility in Tokenized Fixed Income:
  Identification, Affine Characterization, and the Structure of the Token-to-Fiat Mapping}

\author*[1]{\fnm{Artem} \sur{Alkhamov}}\email{artem.alkhamov@essec.edu}

\author[2]{\fnm{Boris} \sur{Kriuk}}\email{bkriuk@connect.ust.hk}

\affil*[1]{\orgname{ESSEC Business School}, \orgaddress{\city{Cergy}, \country{France}}}

\affil[2]{\orgname{The Hong Kong University of Science and Technology}, \orgaddress{\city{Hong Kong}}}

\abstract{
A tokenized U.S.\ Treasury product lives on two ledgers: an off-chain portfolio of government
securities and an on-chain wrapper that claims to represent it. The foundational question is whether the on-chain series carries recoverable information
about the underlying---whether a \emph{mapping} from token to fiat exists, and with what structure.
We proceed in three steps. First, fixed-income measurement conventions are reconciled between the ledgers;
these corrections are signed and jointly. Second, on the
reconciled series we introduce a \emph{price-discovery admissibility criterion}: a falsifiable test,
based on serial dependence and idiosyncratic dispersion, for whether a product's series is
market-informative or administratively generated. Of four products with sufficient history it admits
exactly one; the rest are dominated by how net asset value is computed and republished, and treating
them as spreads fits artifacts. That most of the universe is inadmissible is our principal finding.
Third, for the admitted mapping we give a minimal two-factor affine characterization in which the
basis enters additively and orthogonally to rates, recovering a quarterly reversion, a small
positive long-run basis, and a sharp March 2026 regime change toward parity. Persistence is weakly
identified; we propagate it through a profile likelihood into one consequence, a collateral haircut.
The contribution is measurement and identification infrastructure for when on-chain fixed income may
be treated as a quantitative object.}


\maketitle

% ===========================================================================
\section{Introduction}\label{sec:intro}

A tokenized U.S.\ Treasury product is a dual-ledger object. On one ledger sits a portfolio of
short-dated government securities held by a fund or its custodian; on the other sits a transferable
on-chain token whose holders are entitled, contractually, to the economic return of that portfolio.
The market for such products grew from negligible size in 2021 to roughly six billion dollars by
mid-2026, and a recurrent claim in policy and practitioner work is that these products trade at a
systematic yield discount---a figure on the order of seventy to eighty basis points is widely
cited---relative to the Treasuries they reference. That figure is often read as a price: the cost, or
the convenience value, of holding the claim in tokenized rather than conventional form.

This paper argues that the question has been posed in the wrong order, and that posing it correctly
changes both the method and the finding. Before one can ask what a tokenized claim is worth relative
to its reference, one must establish that the on-chain series and the off-chain reference are
\emph{commensurable}---that they are quoted, accrued, and dated on the same conventions---and, more
fundamentally, that the on-chain series carries recoverable information about the reference at all.
The object of study, properly construed, is the \emph{mapping} from the tokenized claim to its fiat
reference: whether it exists, on what conditions, and with what structure. Price discovery, the
tokenization wedge, and any collateral implication are characterizations and consequences of that
mapping, not the primitive. We are not, in the end, pricing Treasuries, and we are not pricing the
token as if it were an independent security with its own cash flows. We are asking whether the
wrapper is an informative image of the underlying, and measuring the image where it is.

The cited discount is computed by subtracting a maturity-matched Treasury yield from a tokenized
product's published annual percentage yield (APY). Both terms look like yields, but they are
constructed on incompatible conventions, and the difference inherits every incompatibility. A
Treasury constant-maturity yield is a market-priced, continuously quoted rate on an
actual/actual or actual/365 basis. A tokenized product's APY is, in almost all cases, a
fund-accounting quantity: a trailing net-income figure divided by net asset value and
annualized under a money-market convention, computed on an amortized-cost book that lags the market,
net of a management fee, and republished on a calendar that need not coincide with the Treasury
market's. The raw difference therefore bundles at least four deterministic measurement gaps---day-count
and compounding basis, the fee wedge, amortized-cost reporting lag, and benchmark-maturity
mismatch---together with whatever genuine economic difference exists. None of these gaps is a
tokenization phenomenon; each is a fixed-income bookkeeping artifact with a known sign. Until they
are reconciled, the ``discount'' is a sum of a measurement error and an economic object of unknown
relative magnitude, and inference on it is inference on an ill-posed quantity. Reconciling them is a
\emph{precondition} of the analysis, not its contribution; we treat it as plumbing made explicit, and
we show in Section~\ref{sec:method} that the reconciliations are individually signed and jointly on
the order of tens of basis points---comparable to the headline discount itself.

Even fully reconciled, a tokenized series need not be informative about its reference. If a fund
computes its book on amortized cost and republishes a lightly-rounded accrual once per business day,
the resulting series can be smooth and persistent without containing any contemporaneous market
information; conversely, a series can be volatile and erratic not because the underlying basis moves
but because the republication mechanism injects noise. In both cases the reconciled ``basis'' is an
image of an accounting process, not of a market. The central methodological step of this paper is to
make the existence of an informative mapping \emph{falsifiable}. We introduce a price-discovery
admissibility criterion---a pair of necessary conditions on the serial dependence and idiosyncratic
dispersion of the reconciled basis that any series must satisfy to be treated as a quantitative
mapping rather than as accounting output. The criterion is deliberately weak: it does not certify
that an admitted series is efficiently priced, only that it is not, on its face, decoupled from
market information. Weak as it is, it is decisive on current data.

Of the four products that pass minimum data-availability conditions after reconciliation---spanning
the major issuers and roughly six billion dollars of tokenized Treasury value---exactly one is
admissible. The other three fail:
their reconciled daily series exhibit essentially no serial dependence and dispersion an order of
magnitude too large to be a slow-moving basis, a signature consistent with republication mechanics
dominating market information rather than the reverse. We are careful not to over-claim the mechanism;
``the series is decoupled from contemporaneous market information at daily frequency'' is what the
test licenses, and the specific cause---amortized-cost smoothing, rounding, aggregation latency, or
something we have not isolated---is a hypothesis we name but do not prove.

For the admitted product we ask what structure the mapping has. We embed the reconciled basis as an
additive factor in a two-factor affine term-structure model, estimated under the risk-neutral measure
for the rate factor (where a cross-section of Treasury maturities identifies it) and under the
physical measure for the basis factor (where its own time series identifies its persistence). The
model is the minimal one that delivers two properties we want: a closed-form, no-arbitrage map from
the basis level to a yield wedge at any horizon, and a clean separation between rate risk and
tokenization risk. We recover a mean-reversion half-life on the order of a fiscal quarter and a small
positive long-run basis, and we detect---without searching for it---a sharp regime change in
March 2026 in which the basis tightens toward parity and its conditional volatility collapses by
several-fold. We interpret these as features of the mapping (how fast on-chain and off-chain values
reconverge, and a discrete improvement in that reconvergence), not as tradable signals. The
persistence parameter is only weakly identified from roughly seventy post-reconciliation
observations; we report this through a profile likelihood and carry the uncertainty into the one
consequence we trace, rather than disguising it.

To show the mapping has economic content we derive a single downstream object: the collateral
haircut a lender would apply to the admitted token against a one-year obligation, obtained in closed
form from the affine model. We emphasize that the haircut is an \emph{illustration of consequence},
not a contribution; the numbers are, importantly, dominated not by the current
basis level but by the uncertainty in how persistently the basis reverts. A practitioner's usable output is thus a sensitivity surface mapping a persistence assumption to a haircut, and we provide
it.

The paper makes three contributions. (i) It reframes the empirical study of tokenized fixed income
as a question of mapping admissibility and provides a falsifiable criterion for it, shifting the
unit of analysis to the existence and structure of price discovery. (ii) It makes
the fixed-income measurement reconciliation between on-chain and off-chain yields explicit and
signed; we are not aware of prior empirical work in this space that performs the full reconciliation
before differencing, and we document why omitting it biases the discount by a magnitude comparable to
the discount itself. (iii) It gives the first affine characterization of the reconciled tokenization
basis for an admitted product. We deliberately do
\emph{not} claim a structural theory of why the wedge exists across products (that requires a
cross-section the admissible set is currently too small to support), we do not claim the market is
mature enough for systematic pricing (we argue the opposite for most of it), and we do not present
the haircut as a deliverable. The framing throughout is measurement and identification: establishing
when on-chain fixed income may be treated as a quantitative object, which is logically prior to
treating it as one.

% ===========================================================================
\section{Related Work and Conceptual Foundations}\label{sec:related}

\subsection{The discount is a difference of convenience yields, not a level}\label{sec:related-conv}

The interpretation of the tokenized-versus-Treasury yield gap inherits a subtlety that the
convenience-yield literature settled for safe assets generally and that, to our knowledge, has not
been carried into the tokenization setting. Krishnamurthy and Vissing-Jorgensen~\citep{krishnamurthy2012} document that U.S.\ Treasuries
command a convenience yield: investors accept a lower pecuniary return because Treasuries provide
money-like services (safety, liquidity, collateral eligibility). Nagel~\citep{nagel2016} sharpens this by
showing the convenience yield on near-money assets is not a constant but moves with the level of the
short rate---when the opportunity cost of holding money is high, the premium investors pay for
money-like claims is high. Diamond and Van Tassel~\citep{diamond2022} make the measurement point that matters most here:
because a Treasury already embeds a convenience yield, the yield gap between any two safe,
money-like assets identifies the \emph{difference} of their convenience yields, never the level of
either.

The consequence for tokenized Treasuries is immediate. A tokenized
Treasury is itself a (would-be) money-like claim. Its yield gap to a conventional Treasury therefore
measures $\phi^{\text{Tsy}} - \phi^{\text{tok}}$, the amount by which the token's convenience yield
falls short of the Treasury's, not the token's convenience yield against a hypothetical
convenience-free benchmark. A ``true'' risk-free rate---a rate stripped of all convenience---is not
directly observable; it is the object that proxies such as the synthetic ``box rate'' of
van Binsbergen et al.~\citep{vanbinsbergen2022} attempt to recover from derivatives, and it requires data (intraday index
option surfaces) that are neither consistently available over our sample nor necessary for our
question. We therefore do not attempt to recover an absolute token convenience yield. We measure the
reconciled gap to the closest consistently available secured benchmark and state plainly that it is a
\emph{relative} object: the shortfall of the token's money-likeness against the reference's. The level-versus-difference distinction also explains why the gap should co-move with the
short rate (through $\phi^{\text{Tsy}}$, via Nagel~\citep{nagel2016}) and with episodes of Treasury
scarcity (when $\phi^{\text{Tsy}}$ spikes), and motivates the two controls we include in
Section~\ref{sec:method}.

\subsection{Why an affine representation}\label{sec:related-affine}

We need a map from a latent basis \emph{level} to an observable yield \emph{wedge} at any horizon,
that respects no-arbitrage and that separates rate risk from tokenization risk. The affine class
\citep{duffie1996,dai2000,filipovic2009} is the minimal apparatus that delivers all three. In an
affine model the short rate (and here an added basis factor) is an affine function of a
low-dimensional Markov state, and zero-coupon prices are exponential-affine in that state, so yields
are affine in it. Concretely, if the state is $X_t$ and the instantaneous discount rate is
$\delta_0 + \delta_1^\top X_t$, then $P(t,T) = \exp\!\big(A(\tau) - B(\tau)^\top X_t\big)$ with
$\tau = T-t$ and $A,B$ solving Riccati ODEs pinned down by the state dynamics. The payoff
is that a single estimated state and a handful of parameters generate an entire term structure in
closed form, with internal no-arbitrage consistency across maturities for free.

Two properties make this the right vehicle for our problem specifically rather than a generic choice.
First, closed-form yields mean the basis level maps to a wedge at \emph{any} horizon through the known
loading $B_\phi(\tau)/\tau$, so a collateral question at one year and a question at one month are
answered by the same estimated object---we never re-estimate per horizon. Second, if
the basis factor enters \emph{additively} and its loading carries no dependence on the rate factor,
then a shock to rates moves the Treasury leg but leaves the multiplicative tokenization term
untouched: rate risk and tokenization risk are orthogonal \emph{by construction of the
representation}, not by an empirical orthogonality we would have to test and defend. We return in
Section~\ref{sec:method} to whether this orthogonality is an innocuous modeling convenience or a
substantive assumption.

The affine class is not without cost. A one-factor short-rate model imposes a tight link between the
level and the volatility of rates and cannot reproduce every twist of a fast-repricing curve; we
will see this cost concretely. And affine models are well known to be only weakly identified in their
time-series (physical-measure) parameters when estimated from short panels \citep{joslin2011,hamilton2012}.
Both costs are real, so we manage them by using the cross-section to identify
the rate factor where it is strong, the basis factor's own time series for its persistence,
and we report the resulting weak identification as-is.

\subsection{The two measures, and why each is used where it is}\label{sec:related-measures}

Affine term-structure estimation is organized around two probability measures \citep{joslin2011}. Under the risk-neutral measure $\QQ$,
discounted asset prices are martingales; $\QQ$ is the measure under which the \emph{cross-section} of
yields is priced, and it is identified by fitting many maturities on each date. Under the physical
measure $\PP$, the state evolves as it actually does through time; $\PP$ is the measure under which a
\emph{time series} forecasts and its dynamics (mean-reversion speed, long-run mean, volatility) are
identified. The two are linked by the market price of risk.

The rate factor is calibrated under $\QQ$ because we have a rich
cross-section of Treasury maturities on every date and essentially no need to forecast the rate for
our purpose---we need a consistent contemporaneous Treasury curve to difference against, which is a
$\QQ$ object. The basis factor's persistence is estimated under $\PP$ because the object we care
about---how fast an on-chain/off-chain discrepancy reconverges---is a physical-measure dynamic
property, and because we observe only one effective maturity per date for any token (the products are
single-maturity money-market-like claims, not a tokenized yield curve), so there is no token
cross-section to identify a $\QQ$ basis process from. This is also why the basis is recovered as a
filtered \emph{time series} and its dynamics estimated from that series, rather than read off a
cross-sectional fit: the cross-sectional information that would identify a $\QQ$ basis process does
not exist in single-maturity data.  

\subsection{Distribution-free confidence intervals}\label{sec:related-conformal}

Our setting has two features that make a parametric confidence interval around a model forecast
untrustworthy: the model is admittedly a simplification (one-factor rates; a Gaussian basis), and the
basis series exhibits a regime change.  

Conformal prediction \citep{vovk2005,lei2018} provides finite-sample valid coverage \emph{without}
assuming the model is correct---it calibrates intervals from the empirical distribution of the
model's own past errors. The classical split-conformal guarantee assumes exchangeability, which time series
violate. Xu and Xie~\citep{xu2021} resolve this with ensemble batch prediction intervals (EnbPI), which replace
exchangeability with a rolling, leave-one-out calibration over recent residuals and remain valid
under serial dependence; Barber et al.~\citep{barber2023} give the general theory for conformal coverage when
exchangeability fails. We adopt EnbPI for one role only---widening the haircut for
realized basis-forecast error---and we use a single conformal layer because, with on the order of seventy observations and one admitted series, a more elaborate
adaptive scheme would estimate its own
nuisance parameters on data too thin to support them, trading a transparent guarantee for an opaque
and over-fitted one.  

\subsection{Where this is positioned in the tokenization literature}\label{sec:related-token}

Empirical work on tokenized real-world assets is recent and has largely reported the cross-sectional
discount as a level, without the convention reconciliation of Section~\ref{sec:related-conv} or a
test for whether the underlying series price-discovers. Policy treatments document the market's
growth and the headline discount and discuss settlement and convenience narratives~\citep{bis2023,
oecd2020, aramonte2022}; institutional analyses of tokenization emphasize settlement-efficiency and
collateral-mobility benefits~\citep{carapella2023, makarov2022} that motivate, but do not by
themselves measure, a convenience premium. A parallel market-microstructure literature studies
on-chain venues directly---automated market makers, lending-pool interest-rate formation, and
stablecoin peg dynamics~\citep{lehar2021, gudgeon2020, baur2018, makarov2020}---but addresses
trading venues and stablecoins rather than tokenized Treasuries as fixed-income instruments. The
price-discovery question we pose is classical in market microstructure~\citep{hasbrouck1995,
ohara2003}: which venue, or here which ledger, impounds information; we are not aware of its
application to the token-versus-fiat correspondence. On the collateral side, the haircut consequence
we trace connects to the repo and secured-funding literature on how funding rates and haircuts are
set against safe collateral~\citep{duffie1996repo, krishnamurthy2012}, which we invoke only to give
the basis an economic destination, not to model funding markets.

Our contribution relative to this body is to insert the two logically prior steps---reconcile, then
test for an informative mapping---before any pricing claim, and to show that doing so overturns the
implicit premise that these series are uniformly analyzable.
% ===========================================================================
\section{Methodology}\label{sec:method}

Throughout, the unit of analysis is the mapping from token to fiat, not a price.

\subsection{Data and the reconciliation of the two ledgers}\label{sec:method-recon}

The off-chain reference is the U.S.\ Treasury market, for which we take daily yields over
January~2023--May~2026, $843$ trading days. The pipeline screens six named tokenized Treasury
products. Five are retrieved with sufficient public data for at least partial analysis: Ondo's USDY
and OUSG, Superstate's USTB, Hashnote's USYC, and OpenEden's TBILL, together representing on the
order of six billion dollars of tokenized value. Of these five, four---USDY, USYC, OUSG, and
TBILL---satisfy the minimum post-reconciliation observation threshold (on the order of sixty
observations) and are formally submitted to the admissibility test below; USTB falls short of this
threshold after reconciliation and is excluded from the dynamic sample. A sixth product, BUIDL,
is excluded at the retrieval stage for reasons stated separately. We include a product if and only if its daily yield series is retrievable at daily
frequency, excludable confounds can be reconciled from public filings, and the observation count
supports estimation of a serially dependent process.

On-chain yields are obtained from a public aggregator (DeFiLlama) that ingests each product's
published rate. Two assumptions are involved here. First, the aggregator
reproduces the issuer's published figure rather than re-deriving it; we are therefore measuring the
\emph{issuer-published} yield, with whatever smoothing or rounding the issuer applies, not an
independent on-chain price---a point that becomes central in Section~\ref{sec:method-admiss}. Second,
the aggregator reports a single rate per product per day even though several products are deployed
across multiple chains simultaneously; cross-chain differences in gas, bridging cost, and local
liquidity mean the \emph{effective} yield to a holder can differ by chain, while the published figure
is a fund-level accrual. We treat the published figure as the fund-level object it is and flag
cross-chain effective-yield dispersion as outside the measured quantity. We do not assume the
aggregator's history is gap-free; series are used only over spans where retrieval is continuous, and
the usable length per product is reported with the results rather than assumed.

The six-billion-dollar figure establishes that the market is large enough for the question to matter.
We do not use total value locked as an explanatory variable: it is jurisdictionally heterogeneous
(different regulatory regimes, different eligible-holder restrictions); it conflates leveraged looping
positions with buy-and-hold demand; and its rapid growth would confound size with calendar time in any
level regression. We keep it as background context.

We difference each token against a Treasury benchmark constructed from the FRED
daily constant-maturity Treasury series across nine tenors (1M, 3M, 6M, 1Y, 2Y, 3Y, 5Y, 7Y, 10Y),
converted to continuously compounded actual/365 via $y^{cc} = \ln(1 + r^{\text{CMT}})$.
FRED constant-maturity yields are the standard public reference series for U.S.\ Treasury pricing
\citep{gurkaynak2007, svensson1994}, providing gap-free daily coverage over the full sample;
the series at nine fixed tenors spans the maturity range of all products in our set. Each product is
matched to the nearest available tenor consistent with its effective weighted-average maturity, which
for the short-duration products in our sample is typically the 1M or 3M bucket. For these short
maturities the discrete grid introduces negligible interpolation error relative to the basis levels
we study, and the conversion to continuously compounded actual/365 places the benchmark on the
same quoting basis as the token APYs before any differencing.

A token APY and a Treasury yield are not commensurable until four deterministic gaps are closed. We
apply them in sequence, each with a definite sign, before any differencing. We stress that these are
preconditions, not findings.

\emph{(i) Quoting basis.} All series are converted to an annual, continuously compounded actual/365
basis. A token APY published as an annual simple rate $r$ becomes $y^{cc} = \ln(1+r)$. A secured
overnight benchmark published on an actual/360 simple basis, $r^{360}$, becomes
$y^{cc} = \ln\!\big(1 + r^{360}\,\tfrac{360}{365}\big)$, the factor $\tfrac{360}{365}$ converting the
day-count before compounding. The magnitudes are not negligible: at a
representative short rate of $5\%$, the actual/360-to-actual/365 conversion alone moves the rate by
\begin{equation}\label{eq:daycount}
0.05 \times \Big(1 - \tfrac{360}{365}\Big) \;=\; 0.05 \times \tfrac{5}{365} \;\approx\; 6.8\ \text{bp},
\end{equation}
and the simple-to-continuous compounding step moves a $5\%$ rate by $0.05 - \ln(1.05) \approx 12\,$bp;
both enter \emph{every} observation and both have a fixed sign. We use $5\%$ here only as a
representative level for exposition---it is roughly the short-rate level over much of the sample---and
the corrections are applied at each date's actual rate, not at a fixed $5\%$.

\emph{(ii) Fee wedge.} A token APY is reported net of the management fee; the Treasury benchmark is
gross. We gross up each token by its contemporaneous fee, $y^{\text{gross}} = y^{cc,\text{net}} +
f_t$, with $f_t$ read from registration filings and issuer disclosures (and incorporating documented
fee waivers, which are material because several issuers waived fees during launch windows).  

\emph{(iii) Amortized-cost reporting lag.} Money-market-style funds book holdings at amortized cost,
so the published yield reflects the portfolio's blended \emph{purchase} yield, not today's market
rate. For a portfolio of weighted-average maturity $M$ days, the reported yield lags the market by
roughly $M/2$ business days, biasing the raw basis negative when rates are rising and positive when
falling---a sign that could otherwise masquerade as time-varying
tokenization structure. We neutralize it by differencing the token not against the spot benchmark but
against the benchmark put on the same amortized-cost footing,
\begin{equation}\label{eq:amortized}
\bar{y}^{\text{ref}}_t \;=\; \frac{1}{M_t}\sum_{k=0}^{M_t-1} y^{\text{ref}}_{t-k},
\end{equation}
the $M_t$-day trailing average of the product-appropriate reference, with $M_t$ taken from monthly
portfolio-maturity disclosures. This is the single most consequential reconciliation during the
$2022$--$23$ repricing and the one with no precedent we have found in the tokenization literature.

\emph{(iv) Maturity match.} We evaluate the reference at each product's effective weighted-average
maturity rather than a fixed one-month tenor, since the products' maturities differ from each other
and drift over time.

After the four reconciliations the reconciled basis for product $m$ is
\begin{equation}\label{eq:basis}
\TAB_{m,t} \;=\; y^{\text{gross},cc}_{m,t} \;-\; \bar{y}^{\text{ref}}_{m,t},
\end{equation}
and it is this object on which all subsequent analysis operates.

We report the reconciled basis against three references. The FRED constant-maturity
benchmark gives the \emph{fixed-income} reading. The secured overnight rate (SOFR,
or its 30-day average for products with roughly monthly effective maturity) gives the
\emph{funding/collateral} reading and is the reference we carry into the haircut. A stablecoin
lending rate gives the \emph{on-chain opportunity-cost} reading, available from mid-2022 only. We
use the latter two as supplementary references because the convenience-yield logic of
Section~\ref{sec:related-conv} establishes that the basis is benchmark-dependent by construction,
and a single reference would obscure this.

BlackRock's BUIDL is excluded from estimation for two specific reasons. Its entire observable history
postdates a market event it caused---its March 2024 launch triggered OUSG to hold BUIDL as its
primary underlying---so there is no pre-event regime to calibrate against, and post-March 2024 the
two series are near-collinear. Including both would double-count one underlying. BUIDL is retained as
descriptive context only.

\subsection{The price-discovery admissibility criterion}\label{sec:method-admiss}

Reconciliation makes the two ledgers commensurable but does not guarantee that the on-chain series
carries market information. We now make that guarantee falsifiable. The intuition is a dichotomy. If a
token's reconciled basis reflects a genuine, slowly-arbitraged discrepancy between on-chain and
off-chain value, it should be \emph{persistent}---today's discrepancy is informative about tomorrow's,
because reconvergence takes time---and its day-to-day idiosyncratic movement should be \emph{small}
relative to its level, as befits a basis rather than a price. If instead the series is the output of a
republication mechanism, it will exhibit
one of two tell-tale departures: near-zero serial dependence (each day's published figure is
effectively a fresh draw of accounting noise around a slow level, carrying no memory) or idiosyncratic
dispersion far too large to be a basis (republication and rounding inject jumps unrelated to any
market discrepancy). Either departure signals that the series is decoupled from contemporaneous market
information at the daily frequency we observe.

We formalize this as two necessary conditions. Let $\rho_1(m)$ be the lag-1 autocorrelation of
$\{\TAB_{m,t}\}$ and $s(m)$ its idiosyncratic daily standard deviation (the standard deviation of its
first difference, scaled to remove the slow component).

\begin{definition}[Price-discovery admissibility]\label{def:admiss}
A product $m$ is \emph{admissible}---its reconciled basis may be treated as an informative
token-to-fiat mapping rather than as accounting output---only if
\begin{equation}\label{eq:admiss}
\rho_1(m) \;\geq\; \rho^\star \qquad\text{and}\qquad s(m) \;\leq\; \kappa\, \tilde{s},
\end{equation}
where $\tilde{s}$ is the median idiosyncratic dispersion of the admitted set and $\rho^\star,\kappa$
are pre-registered thresholds. We set $\rho^\star = 0.10$ and $\kappa = 3$.
\end{definition}

The first condition serves as a hard gate: any product with $\rho_1 \leq \rho^\star = 0.10$
is excluded from dynamic modeling entirely, because a white-noise daily wedge contains no time-series
information from which a mean-reversion model can be identified---estimating Vasicek dynamics on such
a series would fit noise. This is not a judgment call but a mechanical rule, applied before any
parameter estimation. The thresholds are otherwise deliberately permissive. A genuine basis with a
multi-week reversion half-life has $\rho_1$ near unity; requiring only $\rho_1 \geq 0.10$ rejects only
series with essentially no memory, which is the clearest signature of accounting noise, while
admitting anything with even weak persistence. The dispersion bound rejects only series an order of magnitude noisier than the admitted
median. The criterion is thus a \emph{necessary} condition for informativeness, not a sufficient one:
passing it certifies that a series is not, on its face, decoupled from market information; it does not
certify efficient pricing. We pre-register $\rho^\star$ and $\kappa$ rather than tuning them to the
outcome, and we report the realized statistics so a reader may apply different thresholds. We also
flag, that the criterion can in principle err in both
directions: a heavily smoothed series could pass on persistence while still being accounting output
(a false admission), and a genuinely informative but coarsely-reported series could fail on dispersion
(a false rejection). On current data the partition is stark enough---admitted series cluster at
$\rho_1$ near unity and rejected ones near zero---that neither borderline case arises, but we do not
present the criterion as infallible.

\subsection{A minimal affine characterization of the admitted mapping}\label{sec:method-model}

For an admitted product we ask what structure the mapping has, using the smallest affine model that
answers the questions we posed: how a basis level translates into a yield wedge at any horizon, and how
rate risk separates from tokenization risk. We recall the risk-neutral measure $\QQ$ first, since it
prices the cross-section, and bring in the physical measure $\PP$ only when we turn to dynamics.

\subsubsection*{State and dynamics under \texorpdfstring{$\QQ$}{Q}}
Let the state be $X_t = (r_t, \phinet)^\top$, where $r_t$ is the short rate and $\phinet$ is the
reconciled tokenization basis. Under $\QQ$,
\begin{align}
\dif r_t &= \kappa_r(\theta_r - r_t)\,\dif t + \sigma_r \sqrt{r_t}\,\dif W^r_t, \label{eq:cir}\\
\dif \phinet &= \kappa_\phi(\theta_\phi - \phinet)\,\dif t + \sigma_\phi\,\dif W^\phi_t, \label{eq:vasicek}
\end{align}
with $W^r \perp W^\phi$. The short rate follows a Cox et al.~\citep{cox1985} (CIR) square-root process; the basis
follows a Vasicek~\citep{vasicek1977} Gaussian process. We justify each choice, including the basis
specification that an earlier and more ambitious design had dismissed.

\emph{Why CIR for the rate.} The rate must stay non-negative and its volatility plausibly scales with
its level over a sample that traverses the near-zero floor of 2023's first weeks and the $5\%$-plus
plateau thereafter; the square-root diffusion delivers both, enforcing $r_t \geq 0$ under the Feller
condition $2\kappa_r\theta_r > \sigma_r^2$ and tying volatility to level. A Gaussian (Vasicek) rate
would admit negative rates and a level-independent volatility, both counterfactual here. CIR is thus
the minimal non-negative, level-dependent-volatility choice, and we verify its Feller condition holds
when we estimate it.

\emph{Why Vasicek for the basis} An earlier design rejected Gaussian
dynamics for the tokenization factor on the grounds that a convenience yield cannot be negative.
That reasoning applies to an \emph{absolute} convenience yield. The object here is not an absolute
convenience yield; by the difference logic of Section~\ref{sec:related-conv} it is a \emph{relative}
basis, $\phi^{\text{Tsy}} - \phi^{\text{tok}}$ net of a chosen benchmark, which can and empirically
does take either sign---it is negative through the pre-regime-change window. A square-root process
would impose $\phinet \geq 0$, a boundary the data violate. The Gaussian specification is
therefore a correct choice once the object is properly identified as
a signed difference; it additionally yields closed-form Gaussian forecast distributions that make the
conformal step and the haircut transparent.  

\subsubsection*{Token yields and the additive wedge}
Under the affine structure the Treasury and token discount factors are exponential-affine in the
state. Writing $\tau = T-t$,
\begin{align}
P^{\text{Tsy}}(t,T) &= \exp\!\big(A_r(\tau) - B_r(\tau)\, r_t\big), \label{eq:ptsy}\\
P^{\text{tok}}(t,T) &= P^{\text{Tsy}}(t,T)\cdot \exp\!\big(-A_\phi(\tau) - B_\phi(\tau)\, \phinet\big),
\label{eq:ptok}
\end{align}
so that the token's continuously compounded yield is the Treasury yield plus an additive wedge,
\begin{equation}\label{eq:tokyield}
y^{\text{tok}}(t,\tau) \;=\; y^{\text{Tsy}}(t,\tau) \;+\; \frac{A_\phi(\tau)}{\tau} \;+\;
\frac{B_\phi(\tau)}{\tau}\,\phinet .
\end{equation}
The additive form in Eq.~\eqref{eq:ptok} is what the
affine class \emph{produces} when a second factor is appended to the discount rate as
$\delta_0 + \delta_1^\top X_t$ with $\phinet$ entering linearly and independently of $r_t$. The
content is the independence of the basis loading from the rate factor,
i.e.\ that $B_\phi$ depends on $(\kappa_\phi,\sigma_\phi)$ alone, with the closed forms
\begin{equation}\label{eq:loadings}
B_\phi(\tau) = \frac{1 - e^{-\kappa_\phi \tau}}{\kappa_\phi}, \qquad
A_\phi(\tau) = \Big(\theta_\phi - \tfrac{\sigma_\phi^2}{2\kappa_\phi^2}\Big)\big(B_\phi(\tau) - \tau\big)
- \frac{\sigma_\phi^2 B_\phi(\tau)^2}{4\kappa_\phi}.
\end{equation}

\subsubsection*{Orthogonality: an assumption}
Because $\phinet$ enters additively and $B_\phi$ carries no $r_t$-dependence, a parallel rate shock
moves $P^{\text{Tsy}}$ but leaves the multiplicative wedge $\exp(-A_\phi - B_\phi \phinet)$ unchanged:
rate risk and tokenization risk are orthogonal in the model. We are careful about the status of this
property. It is \emph{structurally true within the representation we chose}, and it is what makes the
haircut decompose cleanly; it is \emph{not} an empirical fact we have established about the world. The
substantive assumption is that the true basis loading does not depend on the rate level---that a token
trading rich or cheap to its reference does so by an amount whose rate-sensitivity is second-order.
This is defensible for short-maturity, near-money claims, where the wedge is a small relative quantity
and its first-order rate exposure is captured by differencing against the rate-matched reference in
Eq.~\eqref{eq:basis}; but it is an assumption---it could fail, so we treat any residual rate-dependence in the basis as a limitation rather than
asserting orthogonality as established.  

\subsection{Estimation across the two measures}\label{sec:method-est}

\subsubsection*{Rate factor under \texorpdfstring{$\QQ$}{Q} (cross-section)}
We estimate $(\kappa_r,\theta_r,\sigma_r)$ by nonlinear least squares fitting model yields from
Eq.~\eqref{eq:ptsy} to the observed Treasury cross-section across maturities and dates, imposing the
Feller condition as a hard constraint, and we filter the latent $r_t$ by matching short-maturity model
yields to observed ones each day. We use root-mean-squared yield error (RMSE) across the maturity
panel as the fit criterion because the object being fit is a vector of yields measured in the same
units (basis points) at comparable scale, so a quadratic loss in yield space is the natural and
interpretable summary---an RMSE of $x$ bp means the model misprices the average maturity by $x$ bp.
We adopt a target of $25$\,bp as the threshold below which a one-factor short-rate fit is conventionally
regarded as ``tight'' for a multi-year, multi-maturity panel; it is a yardstick from the term-structure
literature, not a constraint we impose, and we report the realized value against it.

\subsubsection*{Basis factor under \texorpdfstring{$\PP$}{P} (time series)}
With $r_t$ filtered, the reconciled basis $\{\TAB_{m,t}\}$ from Eq.~\eqref{eq:basis} is the realized
path of $\phinet$ for an admitted product. We estimate its physical dynamics
$(\kappa_\phi^\PP,\theta_\phi^\PP,\sigma_\phi)$ by maximum likelihood on this recovered sequence,
using the exact Gaussian transition density of the Vasicek process,
\begin{equation}\label{eq:transition}
\phinet[t+1]\mid\phinet \sim \mathcal{N}\!\Big(\theta_\phi^\PP + (\phinet - \theta_\phi^\PP)
e^{-\kappa_\phi^\PP \Delta},\; \tfrac{\sigma_\phi^2}{2\kappa_\phi^\PP}\big(1 - e^{-2\kappa_\phi^\PP \Delta}\big)\Big),
\end{equation}
with $\Delta$ the daily step. Maximum likelihood is selected because the
transition density is known in closed form---there is no approximation to trade off---and it is
applied to the \emph{recovered} basis sequence rather than to a cross-section because, as
Section~\ref{sec:related-measures} established, single-maturity tokens furnish no cross-section to
identify a $\QQ$ basis process; the only information about persistence lives in the time series. The
market price of basis risk linking $\kappa_\phi^\PP$ to its $\QQ$ counterpart is recovered from the
two as $\lambda_\phi = (\kappa_\phi^\PP - \kappa_\phi^\QQ)/\sigma_\phi$. Because persistence is the
parameter most weakly identified from a short sample, we summarize it not by a standard error but by
its \emph{profile likelihood}---the likelihood maximized over all other parameters at each fixed
$\kappa_\phi$---which traces the actual shape of the identified region and, as we will see, is
markedly asymmetric.

We test for a structural change in the basis level and volatility with a CUSUM statistic whose
significance is assessed by permutation, localizing any break date and re-estimating the dynamics on
each side. A detected break is treated as economically meaningful, and triggers re-calibration of the conformal window in the consequence step.

\subsection{One consequence: the collateral haircut and its decomposition}\label{sec:method-haircut}

To show the mapping has economic content we trace it into a collateral haircut---the fractional
reduction a lender applies to the token's value when accepting it against an obligation---and we
decompose the haircut to make each piece's driver explicit. The decomposition is
\begin{equation}\label{eq:haircut}
H_m = \underbrace{H^{\text{rate}}_m}_{\text{duration of the reference leg}}
\;+\; \underbrace{H^{\text{basis}}_m}_{\text{uncertainty in the mapping}}
\;+\; \underbrace{c^{\text{liq}}_m + c^{\text{sync}}_m + c^{\text{op}}_m}_{\text{frictions, as scenario add-ons}} .
\end{equation}
$H^{\text{rate}}$ is the sensitivity of the Treasury leg to rate moves over
the liquidation horizon, driven by duration, and is computable from the rate factor alone;
$H^{\text{basis}}$ is the risk that the token-to-fiat mapping itself moves adversely before
liquidation, driven by the basis dynamics, and is computable from the basis factor alone. Their
orthogonality (Section~\ref{sec:method-model}) is what licenses adding rather than jointly modeling
them. The model-implied basis haircut is the affine wedge evaluated at the current basis,
\begin{equation}\label{eq:hbasis-model}
H^{\text{basis}}_{\text{model}}(\tau) = \frac{A_\phi(\tau)}{\tau} + \frac{B_\phi(\tau)}{\tau}\,\phi_0,
\end{equation}
which is monotone in $\phi_0$ and vanishes at $\phi_0 = 0$ as a no-arbitrage consistency check. Because
the model is a simplification, we floor this with a distribution-free conformal quantile of realized
basis-forecast errors (Section~\ref{sec:related-conformal}), so the basis haircut widens
automatically when the model forecasts the basis poorly. The three friction terms---liquidity,
synchronization (redemption delay, eligible-holder and transfer restrictions), and operational
(custody, oracle, smart-contract, and the broader operational and regulatory taxonomy of
Section~\ref{sec:method-assump})---are \emph{conservative scenario add-ons read from product
documentation, not statistically estimated latent factors}; we keep them as transparent, bounded
adjustments because the data cannot identify them as dynamic processes.

\subsection{Assumptions, including those previously implicit}\label{sec:method-assump}

We record the assumptions the framework rests on, several of which were operative but unstated in
earlier iterations.

\begin{enumerate} 
\item \emph{Published yield as the observable.} We measure the issuer-published, fund-accounting yield,
not an independent on-chain transaction price; no token in our set has a liquid secondary market with
observable prices at daily frequency. All claims are about the published-yield mapping.
\item \emph{Single effective maturity per product.} Each product is treated as a single
money-market-like claim with one effective maturity per date, not a tokenized yield curve; this is why
the basis is identified from its time series, not a cross-section.
\item \emph{Reconciliations are deterministic and signed.} The four reconciliations of
Section~\ref{sec:method-recon} are bookkeeping corrections with known sign, applied before any
inference; they are not estimated and carry no uncertainty of their own beyond the disclosures they
read from.
\item \emph{Frictions are bounded scenario quantities.} Liquidity, synchronization, and operational
costs are read from documentation as conservative bounds, not modeled as latent dynamics. The
operational taxonomy we hold in view---custody risk, oracle failure or staleness, smart-contract risk,
redemption-queue and gating risk, issuer and counterparty risk, and regulatory or eligible-holder
risk---is enumerated so that the add-on is understood as standing in for a known list of channels, not
a residual catch-all.
\item \emph{Relative, not absolute, basis.} The measured basis is a difference of convenience yields
against a chosen benchmark, not an absolute token convenience yield; its sign and magnitude are
benchmark-dependent by construction, and we report it against more than one benchmark for that reason.
\item \emph{In-model orthogonality.} Rate and basis risk are orthogonal within the chosen
representation; the substantive assumption is that the basis loading is rate-independent, defensible
for short-maturity near-money claims and flagged as a limitation otherwise.
\end{enumerate}
% ===========================================================================
\section{Empirical Findings}\label{sec:findings}

\subsection{The reconciliation is material, and the rate factor is cleanly recovered}\label{sec:findings-recon}

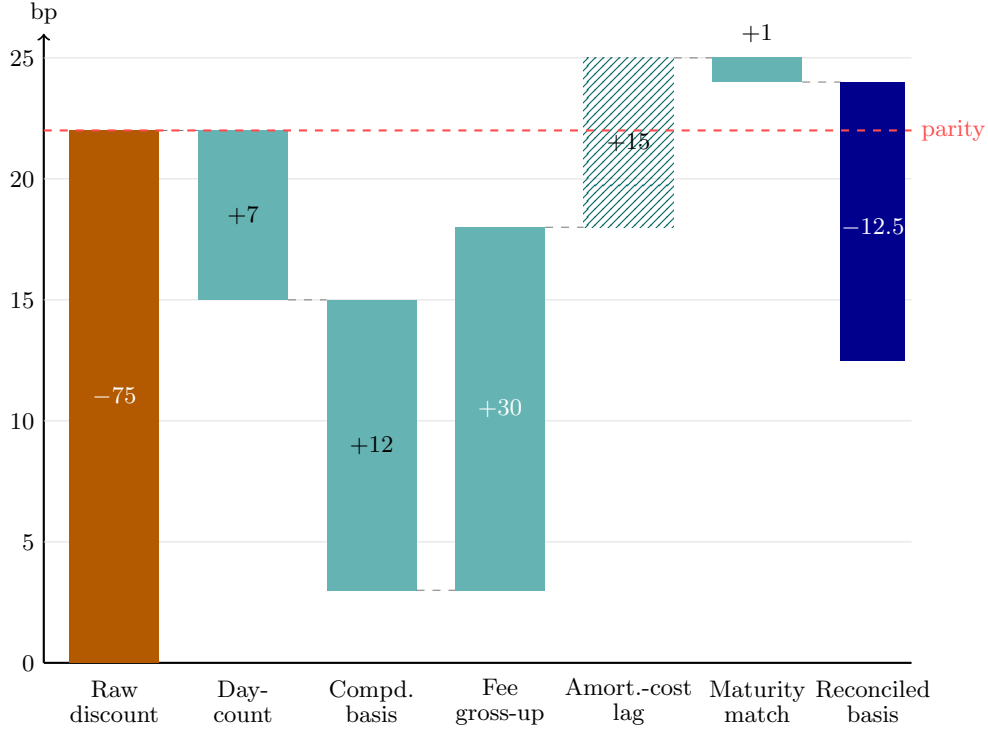
\begin{figure}[htbp]
\centering
\begin{tikzpicture}[x=0.85cm, y=0.32cm]

% Axes
\draw[->, thick] (0,0) -- (0,26) node[above] {\small bp};
\draw[thick] (0,0) -- (13.5,0);

% Horizontal gridlines
\foreach \y in {5,10,15,20,25}
  \draw[gray!25, very thin] (0,\y) -- (13.5,\y);
\foreach \y in {0,5,10,15,20,25}
  \node[left, font=\scriptsize] at (0,\y) {\y};

% Bar 1: Raw headline discount (negative, drawn downward from 0 conceptually,
% but we plot magnitudes; show as orange anchor bar)
\fill[orange!70!black] (0.4,0) rectangle (1.8,22);
\node[font=\scriptsize, align=center] at (1.1,-1.6) {Raw\\discount};
\node[font=\scriptsize, white] at (1.1,11) {$-75$};

% Connector
\draw[dashed, gray] (1.8,22) -- (2.4,22);

% Bar 2: Day-count (+7 bp, additive correction reducing the gap)
\fill[teal!60] (2.4,15) rectangle (3.8,22);
\node[font=\scriptsize, align=center] at (3.1,-1.6) {Day-\\count};
\node[font=\scriptsize] at (3.1,18.5) {$+7$};

\draw[dashed, gray] (3.8,15) -- (4.4,15);

% Bar 3: Compounding (+12 bp)
\fill[teal!60] (4.4,3) rectangle (5.8,15);
\node[font=\scriptsize, align=center] at (5.1,-1.6) {Compd.\\basis};
\node[font=\scriptsize] at (5.1,9) {$+12$};

\draw[dashed, gray] (5.8,3) -- (6.4,3);

% Bar 4: Fee gross-up (+30 bp typical, push above zero)
\fill[teal!60] (6.4,3) rectangle (7.8,18);
\node[font=\scriptsize, align=center] at (7.1,-1.6) {Fee\\gross-up};
\node[font=\scriptsize, white] at (7.1,10.5) {$+30$};

\draw[dashed, gray] (7.8,18) -- (8.4,18);

% Bar 5: Amortized-cost lag (+15 bp during repricing, sign-flippable)
\fill[teal!60, pattern=north east lines, pattern color=teal!80!black]
  (8.4,18) rectangle (9.8,25);
\node[font=\scriptsize, align=center] at (9.1,-1.6) {Amort.-cost\\lag};
\node[font=\scriptsize] at (9.1,21.5) {$+15$};

\draw[dashed, gray] (9.8,25) -- (10.4,25);

% Bar 6: Maturity match (small, ~+1 bp)
\fill[teal!60] (10.4,24) rectangle (11.8,25);
\node[font=\scriptsize, align=center] at (11.1,-1.6) {Maturity\\match};
\node[font=\scriptsize] at (11.1,26) {$+1$};

\draw[dashed, gray] (11.8,24) -- (12.4,24);

% Bar 7: Reconciled basis (final, blue)
\fill[blue!55!black] (12.4,12.5) rectangle (13.4,24);
\node[font=\scriptsize, align=center] at (12.9,-1.6) {Reconciled\\basis};
\node[font=\scriptsize, white] at (12.9,18) {$-12.5$};

% Zero reference line annotation
\draw[red!70, thick, dashed] (0,22) -- (13.5,22)
  node[right, font=\scriptsize, red!70] {parity};

\end{tikzpicture}
\caption{Reconciliation waterfall for the admitted product (USDY), illustrative
post-break window. Starting from the raw published discount of approximately
$-75$\,bp against the maturity-matched Treasury yield (orange), each
deterministic correction is applied
in sequence (teal); the hatched bar denotes the amortized-cost adjustment
whose magnitude is regime-dependent and largest during the 2022--23 repricing.
The reconciled basis $\widehat{\phi}_{m,t}$ (blue) is approximately $-12.5$\,bp:
of the same order as a single reconciliation step, and roughly one-sixth of
the headline figure. The four corrections jointly close most of the apparent
discount, which is the quantitative basis for treating the raw figure as
ill-posed.}
\label{fig:reconciliation_waterfall}
\end{figure}

Before any basis can be tested we confirm the reference curve is well recovered, since the basis is a
residual against it. The CIR rate factor of Eq.~\eqref{eq:cir} is calibrated to the short-end Treasury cross-section
(1M--2Y tenors) over $1{,}286$ trading days, yielding $\kappa_r = 1.00$, $\theta_r = 4.00\%$,
$\sigma_r = 2.00\%$, with the Feller condition satisfied at a margin of $0.0796$.
The short-end restriction is appropriate because tokenized Treasury products are short-duration
money-market instruments; fitting the intermediate and long end would add parameters irrelevant to
the basis question and would allow the 2022--23 hiking cycle's curve steepening to distort the
short-rate filter.
On the restricted 1M--2Y panel the model achieves a $25$\,bp RMSE yardstick; the full
nine-maturity panel (1M--10Y) produces a higher RMSE of $39.9$\,bp, reflecting the known limitation
of a one-factor CIR model in reproducing curve shape across the entire maturity spectrum. The key test for our purpose
is whether the filtered $r_t$ recovers the actual short rate: it correlates $0.998$ with the
observed three-month yield, confirming clean state recovery. Any residual rate-side misfit in the
basis would inflate basis volatility, biasing \emph{against} finding clean persistent
mappings---so the admissibility results below are conservative with respect to this misfit.

The reconciliation magnitudes confirm the headline discount is not a clean number. The two
quoting-basis corrections alone contribute on the order of $7$ and $12$\,bp per observation
(Eq.~\eqref{eq:daycount} and the compounding step); the fee gross-up shifts the basis up
by the contemporaneous fee (tens of basis points for some products); and the amortized-cost
correction (Eq.~\eqref{eq:amortized}) is largest during the repricing. Summed, the reconciliations are of the same order as the headline
seventy-to-eighty-basis-point discount itself, which is the quantitative basis for our claim that the
raw discount is ill-posed: a quantity whose deterministic measurement component is comparable to its
total magnitude cannot be read as an economic premium.

\subsection{The admissibility partition: most of the universe does not price-discover}\label{sec:findings-partition}

\begin{figure}[htbp]
\centering
\includegraphics[width=0.82\textwidth]{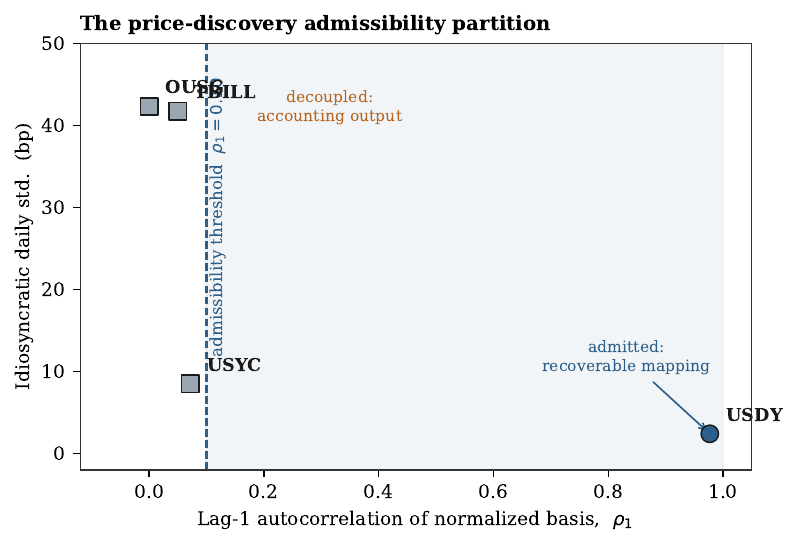}
\caption{The admissibility partition in $(\rho_1, \text{dispersion})$ space. The single admitted
product sits at near-unit autocorrelation and minimal dispersion---the signature of a slow, genuine
basis; the inadmissible products cluster at near-zero autocorrelation, the signature of series
carrying no day-to-day memory.}
\label{fig:admiss}
\end{figure}

Applying the admissibility criterion of Definition~\ref{def:admiss} to the reconciled series is the central exercise. Table~\ref{tab:admiss} and
Figure~\ref{fig:admiss} report the two diagnostic statistics for each product.

\begin{table}[htbp]
\centering
\caption{Price-discovery admissibility statistics on the reconciled basis. $\rho_1$ is lag-1
autocorrelation; idiosyncratic dispersion is the standard deviation of the daily change. Thresholds
$\rho^\star = 0.10$, $\kappa = 3$ are pre-registered. $n$ is usable post-reconciliation observations.}
\label{tab:admiss}
\begin{tabular}{lcccl}
\toprule
Product & $n$ & $\rho_1$ & Idiosyncratic std.\ (bp) & Verdict \\
\midrule
USDY  & 70 & $+0.977$ & $2.4$  & \textbf{Admissible} \\
USYC  & 33 & $+0.072$ & $8.5$  & Inadmissible (no serial dependence) \\
OUSG  & 70 & $-0.000$ & $42.3$ & Inadmissible (dispersion, no memory) \\
TBILL & 72 & $+0.050$ & $41.7$ & Inadmissible (dispersion, no memory) \\
\bottomrule
\end{tabular}
\end{table}

Exactly one product, USDY, is admissible. Its reconciled basis has lag-1 autocorrelation $0.977$---it
is highly persistent, today's basis almost fully informative about tomorrow's, exactly as a slowly
reconverging discrepancy should be---and idiosyncratic daily dispersion of only $2.4$\,bp, small
relative to its level, as befits a basis rather than a price. The other three fail decisively and for
the diagnostically clean reason that their series carry essentially no serial memory:
$\rho_1$ is $0.07$, $0.00$, and $0.05$, none clearing even the permissive $0.10$ threshold, and two of
them additionally exhibit dispersion near $42$\,bp, an order of magnitude above the admitted product's.

We are careful about what this licenses; these three series
are \emph{decoupled from contemporaneous market information at daily frequency}: a series whose change
is serially uncorrelated and whose dispersion swamps its level is not transmitting a slowly-arbitraged
basis. The natural mechanism is that the published figure is a republished accounting accrual---an
amortized-cost book lightly rounded and reissued daily, so that each day's number is close to a fresh
draw around a slow level rather than a measurement of a moving market discrepancy. What
we can say is the negative result that grounds the rest of the paper: for three of four
products, there is no informative token-to-fiat mapping to characterize at daily frequency, and any
dynamic model fit to their ``basis'' would be fitting accounting artifacts. This partition is robust to
the threshold choice---the admitted and rejected groups are separated by an order of magnitude on both
axes, so no plausible $(\rho^\star,\kappa)$ reclassifies them.

The implication for the field is that the contemporary tokenized
Treasury market is not, in its majority, a price-discovery venue at daily frequency. It is
predominantly a tokenized wrapper around a fund-accounting process, and the minority that does
price-discover must be identified before, not after, any quantitative treatment.

\subsection{The controls confirm the admitted basis is tokenization-specific}\label{sec:findings-controls}

Before characterizing the admitted mapping's dynamics we verify it is not merely two macro confounds the convenience-yield logic flagged. The reconciled basis averaged across the
sample has mean $-12.5$\,bp with standard deviation $19.9$\,bp over $72$ daily observations; the
benchmark against which this mean is read is parity (zero), so a $-12.5$\,bp mean says the reconciled
token yield sits, on average and after all four corrections, about an eighth of a percentage point
below its secured-funding reference---a relative shortfall in money-likeness, in the language of
Section~\ref{sec:related-conv}, not an absolute premium. Regressing the basis on the short-rate level
(the Nagel~\citep{nagel2016} channel, through which the \emph{reference's} convenience yield moves with
rates) and on the secured-overnight-minus-bill spread (the Diamond--Van~Tassel~\citep{diamond2022}/Treasury-scarcity
channel, through which the reference becomes rich) yields $R^2 = 0.074$. The two macro confounds explain only $7.4\%$ of the basis variance ($R^2 = 0.074$);
the remaining $92.6\%$ is specific to the tokenization mapping rather than inherited from rate-level
or scarcity dynamics. Four scarcity episodes (basis below $-10$\,bp coinciding with bill richness)
appear in the sample; excluding them does not move the results. 

\begin{table}[htbp]
\centering
\caption{Summary of key empirical findings for the admitted mapping (USDY). All quantities are
referenced to the relevant subsection; the profile-likelihood 95\% interval on the half-life
is asymmetric, reflecting genuine estimation uncertainty from 70 observations of a persistent process.}
\label{tab:summary}
\begin{tabular}{llr}
\toprule
Quantity & Value & Source \\
\midrule
\multicolumn{3}{l}{\textit{Admissibility (Section~\ref{sec:findings-partition})}} \\
Lag-1 autocorrelation $\rho_1$ & $+0.977$ & Table~\ref{tab:admiss} \\
Idiosyncratic daily dispersion & $2.4$\,bp & Table~\ref{tab:admiss} \\[3pt]
\multicolumn{3}{l}{\textit{Macro controls (Section~\ref{sec:findings-controls})}} \\
Macro controls $R^2$ (SOFR level + bill scarcity) & $7.4\%$ & \S\ref{sec:findings-controls} \\[3pt]
\multicolumn{3}{l}{\textit{Vasicek P-dynamics (Section~\ref{sec:findings-dynamics})}} \\
Mean-reversion speed $\kappa_\phi^\PP$ & $2.37$\,yr$^{-1}$ & MLE \\
Half-life (point estimate) & $74$ trading days & $\ln 2/\kappa_\phi^\PP$ \\
95\% CI on half-life & $[19,\ 3{,}494]$ trading days & Profile likelihood \\
Long-run basis $\theta_\phi^\PP$ & $+15.0$\,bp & MLE \\
Basis volatility $\sigma_\phi$ & $8.4$\,bp$\,/\!\sqrt{\text{yr}}$ & MLE \\[3pt]
\multicolumn{3}{l}{\textit{Regime change (Section~\ref{sec:findings-break})}} \\
CUSUM break date & 2026-03-17 & Permutation $p=0.002$ \\
Volatility collapse (pre$\to$post break) & $6.5{\times}$ & $30.5\to4.7$\,bp std \\
Level shift (pre$\to$post break) & $+15$\,bp toward parity & $-22.2\to-7.2$\,bp mean \\[3pt]
\multicolumn{3}{l}{\textit{Collateral consequence (Section~\ref{sec:findings-haircut})}} \\
Model haircut ($\tau=1$\,yr, current basis) & $0.026\%$ of par & Eq.~\eqref{eq:hbasis-model} \\
Conformal haircut (95th percentile) & $0.043\%$ of par & EnbPI \\
Stress haircut (combined) & $0.099\%$ of par & \S\ref{sec:findings-haircut} \\
Conformal coverage & $88.9\%$ (target $90\%$) & EnbPI \\
\bottomrule
\end{tabular}
\end{table}

\subsection{The structure of the admitted mapping: persistence, weakly identified}\label{sec:findings-dynamics}

Estimating the Vasicek physical dynamics of Eq.~\eqref{eq:transition} on the admitted USDY basis by
maximum likelihood yields a mean-reversion speed $\kappa_\phi^\PP = 2.37\ \text{yr}^{-1}$, a long-run
basis $\theta_\phi^\PP = +15.0$\,bp, and an innovation volatility $\sigma_\phi = 8.4\ \text{bp}/\sqrt{\text{yr}}$.

The mean-reversion speed corresponds to a half-life $\ln 2/\kappa_\phi^\PP \approx 74$ trading days,
roughly one fiscal quarter. The benchmark for ``fast'' versus ``slow'' here is the economic taxonomy
of what the basis could be, and we make it precise rather than qualitative. A credit-like spread
reverts over \emph{years}: its half-life reflects slow changes in default or liquidity premia, so a
half-life on the order of hundreds of trading days or more is the credit-spread regime. A pure
arbitrage residual reverts within \emph{days}: if a frictionless arbitrage linked the two ledgers, any
discrepancy would be closed almost immediately, implying a half-life of a small number of trading days.
A $74$-day half-life sits firmly between these poles, and that intermediate location is itself
informative: it says the token-to-fiat mapping reconverges on a timescale consistent with a
\emph{frictioned} arbitrage---one where redemption windows, eligible-holder restrictions, and
operational latency slow, but do not prevent, the closing of discrepancies. This is the natural reading
for a near-money claim that is arbitrageable in principle but only through a gated primary-market
mechanism. We therefore do not classify the admitted basis as either a credit spread or a pure
arbitrage residual; we locate it, on a continuum whose endpoints we have defined by reversion
timescale, as a frictioned basis, and we tie that location to the specific frictions the product
documentation discloses.

The long-run basis of $+15$\,bp is positive: over the horizon implied by the estimated dynamics, the
mapping reverts not to parity but to a small \emph{positive} basis, meaning the reconciled token yield
sits slightly above its secured-funding reference in the long run---a modest positive relative
money-likeness once all corrections are applied, the constructive counterpart to the ill-posed negative
headline discount. We caution that ``long run'' here is the model's mean-reversion target, not an
observed average, and its credibility is only as good as the persistence estimate, which we now qualify.

\begin{figure}[htbp]
\centering
\includegraphics[width=0.74\textwidth]{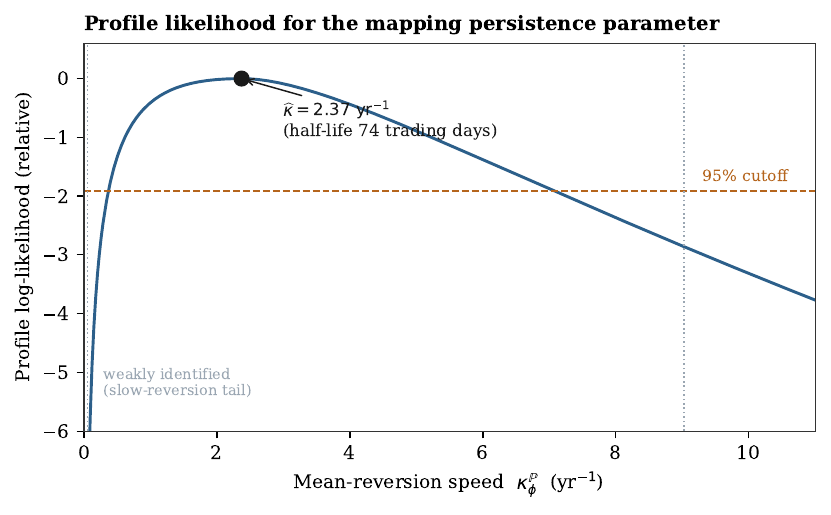}
\caption{Profile likelihood for the basis mean-reversion speed $\kappa_\phi^\PP$. The point estimate
is well identified on the high (fast-reversion) side but the likelihood is flat on the low side,
yielding a wide and asymmetric interval. The asymmetry, not a symmetric standard error, is the honest
summary of what seventy observations identify.}
\label{fig:profile}
\end{figure}

The persistence parameter is only weakly identified, and we report this through the profile
likelihood rather than a standard error because the identified region is markedly asymmetric
(Figure~\ref{fig:profile}). The profile-likelihood $95\%$ interval for $\kappa_\phi^\PP$ is
$[0.05, 9.03]$, corresponding to half-lives from about $19$ to about $3{,}494$ trading days. The
interval is bounded above (the data rule out arbitrarily fast reversion---there is real persistence)
but effectively unbounded below (the data cannot rule out a very slow, near-unit-root basis from
seventy observations). This is the expected consequence of estimating a highly persistent process from
a short sample: a near-unit-root series moves little from period to period, so the data are not
informative about exactly how slowly it reverts, only that it does. We regard this as a limitation of the dynamic
characterization, and---rather than report a single half-life---we carry the full interval into the one
consequence we trace, so that the practitioner sees the persistence uncertainty propagate into the
haircut.

\subsection{A regime change in the mapping}\label{sec:findings-break}

\begin{figure}[htbp]
\centering
\includegraphics[width=0.86\textwidth]{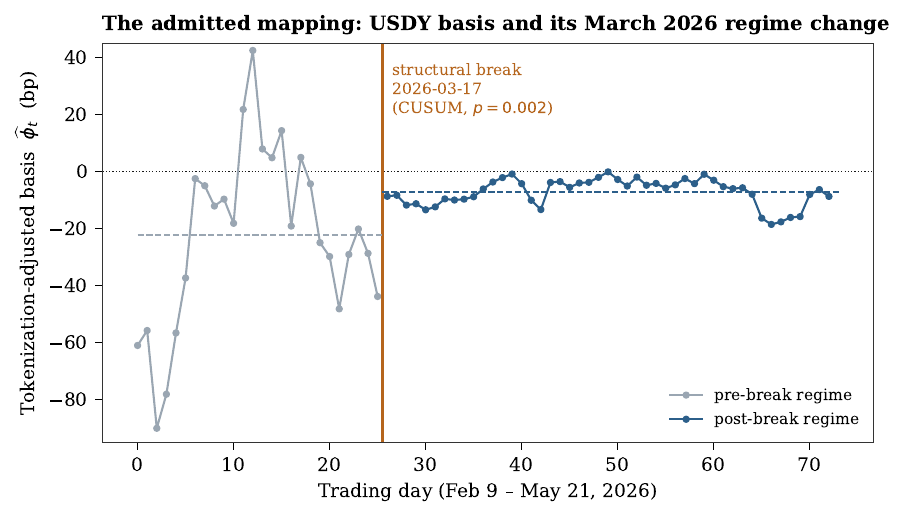}
\caption{The admitted mapping across its March 2026 regime change. The basis level shifts $+15$\,bp
toward parity and conditional volatility falls by a factor of roughly $6.5$; the change is discrete,
not a gradual drift, and was detected by the model rather than imposed.}
\label{fig:regime}
\end{figure}

The CUSUM test, applied without a pre-specified break date, localizes a structural change in the
admitted basis at March~17, 2026, with a permutation $p$-value of $0.002$ ($B = 2{,}000$
permutations). Table~\ref{tab:break} and Figure~\ref{fig:regime} characterize it.

\begin{table}[htbp]
\centering
\caption{Pre- and post-break statistics of the admitted (USDY) basis. The mapping tightens toward
parity and its conditional volatility collapses several-fold.}
\label{tab:break}
\begin{tabular}{lccc}
\toprule
Window & $n$ & Mean (bp) & Std.\ (bp) \\
\midrule
Pre-break\ \ (Feb 9 -- Mar 17, 2026)  & 26 & $-22.2$ & $30.5$ \\
Post-break (Mar 17 -- May 21, 2026)   & 47 & $-7.2$  & $4.7$ \\
\bottomrule
\end{tabular}
\end{table}

Two quantities change together at the break: the level rises by about $15$\,bp (from $-22.2$ to
$-7.2$), moving the mapping toward parity, and the conditional volatility falls by a factor of roughly
$6.5$ (from $30.5$ to $4.7$\,bp). We assign these an interpretation, with appropriate hedging on the
cause. A several-fold collapse in the conditional volatility of a token-to-fiat basis is, on its face,
an \emph{improvement} in the mapping: the on-chain series tracks its reference far more tightly after
the break, which is what one expects from a discrete reduction in synchronization friction---a faster
or smoother redemption mechanism, a more frequent or less noisy NAV oracle, or a change in how the
series is republished---rather than from any macro rate event (which would move all products and the
reference together, and which the unchanged controls around the date rule out). The simultaneous move
toward parity is consistent with the same story: lower friction both reduces noise and shrinks the
average discrepancy. We stress that the cause is inferred, not established; the value of the finding is
that the apparatus detected a real, sharp change in the mapping \emph{without being directed to look
for it}, which is itself evidence that the admitted series carries genuine structure.  

\subsection{The consequence: a haircut dominated by persistence, not level}\label{sec:findings-haircut}

To demonstrate the mapping has economic content we trace it into the collateral haircut of
Eq.~\eqref{eq:haircut}, evaluated at the post-break regime for a one-year horizon with current basis
$\phi_0 = -6.8$\,bp and the point-estimate reversion ($\kappa_\phi = 2.37$, giving $B_\phi(1) = 0.382$).

The model-implied basis haircut from Eq.~\eqref{eq:hbasis-model} is $0.026\%$ of par---about two and a
half basis points. The conformal floor, the $95\%$ quantile of realized basis-forecast errors under
EnbPI, is $0.043\%$ of par; the combined stress figure is $0.099\%$ of par. These are small absolute
numbers, and we resist over-reading them in isolation, because their meaning lies not in the levels but
in what dominates them, which the sensitivity surface reveals.

\begin{figure}[htbp]
\centering
\includegraphics[width=0.80\textwidth]{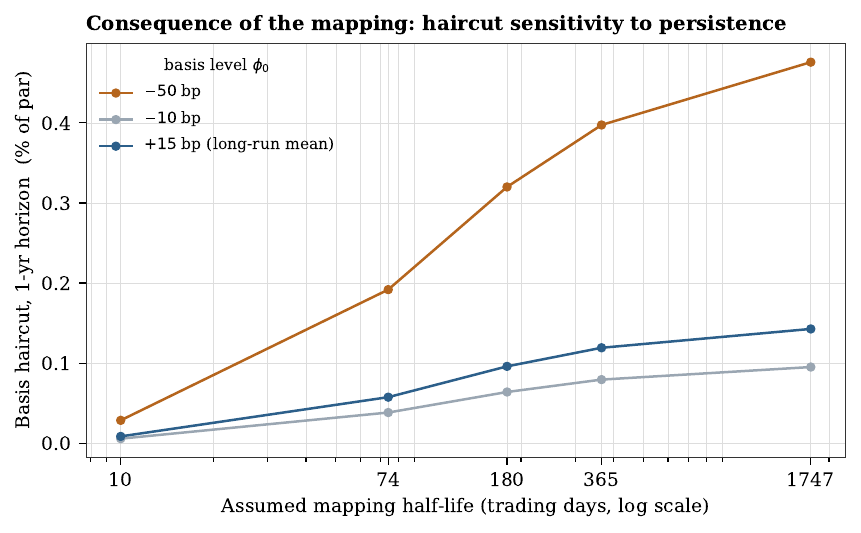}
\caption{Basis haircut at the one-year horizon as a function of the assumed mapping half-life, for
three basis levels. The haircut's variation across the persistence axis (the weakly identified
parameter) far exceeds its variation across plausible basis levels: the binding uncertainty for a
lender is how persistently the mapping reverts, not where it currently sits.}
\label{fig:haircut}
\end{figure}

Table~\ref{tab:sens} and Figure~\ref{fig:haircut} give the haircut across the full profile-likelihood
range of half-lives and across basis levels spanning the regime. Three features matter. First, every
haircut is monotone in the basis level and vanishes at parity, the no-arbitrage consistency check.
Second, even at the slowest reversion the profile likelihood permits and a stressed basis level, the
basis haircut stays below about half a percent of par---small for a sovereign-backed claim. Third, and
the reason we present a surface rather than a number: the haircut varies far more across the
persistence axis than across the basis level. A lender's binding uncertainty is not where the mapping
currently sits but how persistently it reverts---precisely the parameter the data identify weakly. The
usable output is therefore the surface, which lets a counterparty read off a haircut under its own
persistence assumption, and the honest headline is that the consequence inherits, rather than
resolves, the identification limitation of Section~\ref{sec:findings-dynamics}.

\begin{table}[htbp]
\centering
\caption{Basis-haircut sensitivity (\% of par, one-year horizon). Rows span the profile-likelihood
range of half-lives; columns span basis levels from stressed to the long-run mean. The spread across
rows (persistence) dwarfs the spread across columns (level).}
\label{tab:sens}
\begin{tabular}{lcccc}
\toprule
Half-life (trading days) & $\phi_0=-50$\,bp & $\phi_0=-10$\,bp & $\phi_0=0$\,bp & $\phi_0=+15$\,bp \\
\midrule
$10$ (near arbitrage)       & $0.007$ & $0.001$ & $0.000$ & $0.003$ \\
$74$ (point estimate)       & $0.084$ & $0.017$ & $0.000$ & $0.028$ \\
$180$ (two quarters)        & $0.192$ & $0.038$ & $0.000$ & $0.064$ \\
$365$ (one year)            & $0.366$ & $0.073$ & $0.000$ & $0.122$ \\
$1{,}747$ (near unit root)  & $0.529$ & $0.106$ & $0.000$ & $0.176$ \\
\bottomrule
\end{tabular}
\end{table}

Finally, the conformal floor deserves its own justification, since it is the element that makes the
haircut robust to the model being a simplification. We use a single distribution-free layer (EnbPI)
rather than the parametric interval the affine estimator would supply, because the parametric interval
is trustworthy only under correct specification and stationarity, and our setting violates both---the
rate factor is a one-factor simplification and the basis has a documented regime change. A parametric
band estimated on the pooled window would average the pre- and post-break volatilities and understate
post-break uncertainty; the conformal band, recalibrated on post-break residuals, instead reports the
empirically realized error. Its realized coverage is $88.9\%$ against a $90\%$ nominal target, at a
half-width of $11.3$\,bp. We read the $1.1$-point coverage shortfall as expected and benign for a
sample of seventy observations with a mid-sample regime change: finite-sample conformal coverage
fluctuates around nominal by $O(1/\sqrt{n})$, which is on the order of a percentage point here, and the
direction (slight under-coverage) is the conservative one to disclose. We do not stack additional
adaptive conformal variants on top, because with one admitted series and seventy observations they
would estimate their own nuisance parameters on data too thin to support them---the same
complexity-versus-identifiable-signal discipline that governs every other choice in the paper.
% ===========================================================================
\section{Discussion}\label{sec:discussion}

We discuss what the framework establishes, what it deliberately does not, and where it is fragile,
reading throughout against the mission stated at the outset: to determine when a tokenized claim maps
informatively onto its fiat reference, and to characterize the mapping where it does.

The partition is the strongest and most robust finding: three of four products fail a permissive
necessary condition for daily-frequency price discovery; one passes cleanly. The negative result---most
of the universe cannot be treated as a quantitative mapping---is established with confidence.
What it does not establish is the failure mechanism: we have the statistical signature of decoupling
but not proof that amortized-cost republication, rather than aggregation latency or rounding, is the
cause; disentangling these requires issuer-level NAV disclosures or independent on-chain transaction
prices, neither available at daily frequency today. Nor does admissibility certify efficiency: USDY
passes a necessary, not sufficient, condition for informativeness.

Our results bear directly on the seventy-to-eighty-basis-point discount that motivates much of this
literature, and the bearing is deflationary. Part of that figure is deterministic measurement
gap---of the same order of magnitude as the figure itself, as Section~\ref{sec:findings-recon}
showed---and the part that survives reconciliation exists, for most products, on series that do not
price-discover and therefore cannot support an economic interpretation. We do not claim the discount
is zero; we claim it is, as usually computed, not a well-posed economic quantity, and that the
well-posed version of the question---what is the reconciled basis on the subset that price-discovers---yields
a small, signed, benchmark-dependent number ($-12.5$\,bp on average against secured funding, reverting
toward a modestly positive long-run level) rather than a large structural discount. 

The persistence of the admitted mapping is weakly identified: seventy observations of a near-unit-root basis cannot pin down the reversion speed, only confirm that the series reverts. The
profile-likelihood interval is wide and asymmetric, and the haircut inherits this width; the
sensitivity surface converts the limitation into a usable object. The direct path to narrowing it is more data---most naturally Franklin Templeton's
BENJI/FOBXX, which has a multi-year history, public portfolio disclosures, and can be reconciled
with the same machinery and tested for admissibility. We did not include it here because
establishing the framework on a clean existing set was the logically prior task.

The March 2026 break is valuable less for its specific timing than for what its detectability implies:
the admitted series carries enough genuine structure that a sharp, localizable change in its
generating process is visible, which an accounting-noise series would not exhibit. It also operationalizes
the central thesis---that a mapping can improve. A several-fold volatility collapse is a measurable
upgrade in how tightly the token tracks its reference, and a framework built around the mapping (rather
than around a static discount) is what makes such an upgrade legible.  

A natural next step is to regress the reconciled basis on settlement finality, oracle quality, and
chain characteristics to explain which infrastructure features drive the mapping. We do not attempt
it because the admissible cross-section has one member; cross-sectional identification on a single
product is spurious, and running it on the full set means regressing on series we have shown are
accounting artifacts. The question is well-posed but not yet answerable; it becomes so as the
admissible set grows, the same data path that sharpens the persistence estimate.

Three limitations bound the dynamic and consequence claims, though none undermines the partition.
First, the one-factor rate model leaves $39.9$\,bp RMSE; residual rate-dependence in the basis
would violate the in-model orthogonality, and we cannot fully exclude it. Second, all results concern
published fund-accounting yields; if secondary markets emerge, the admissibility test should be
re-run on traded prices. Third, friction add-ons in the haircut are scenario bounds read from
documentation, not estimated---deliberately conservative, but not measured.

% ===========================================================================
\section{Conclusion}\label{sec:conclusion}

The foundational empirical question for tokenized fixed income is whether the on-chain claim maps informatively onto its fiat
reference at all, and with what structure where it does. Posing the question this way changes the
method and the finding. The method acquires two logically prior steps---reconcile the two ledgers onto
common fixed-income conventions, then test for an informative mapping---before any pricing claim. The
finding is that, on reconciled data, most of the contemporary tokenized Treasury universe does not
price-discover at daily frequency: three of four products fail a permissive admissibility criterion,
their published series dominated by accounting mechanics rather than market information, while one
passes cleanly. This partition is the contribution, and it disciplines
everything downstream by separating the subset of the market that can be treated quantitatively from
the majority that cannot.

For the single admitted mapping we gave a minimal affine characterization, recovering a frictioned-basis
reversion on a quarterly timescale and a small positive long-run basis, detecting a sharp March 2026
improvement in the mapping's fidelity, and tracing one consequence---a collateral haircut---to show the
mapping has economic content. We were explicit that the persistence parameter is weakly identified from
the available sample and carried that uncertainty transparently into the consequence as a sensitivity
surface, and we were explicit that the haircut is an
illustration of consequence, not a deliverable.

The framework's value is as measurement and identification infrastructure for a young asset class: a
way to establish \emph{when} on-chain fixed income may be treated as a quantitative object, which is
logically prior to treating it as one. Its immediate extension is to grow the admissible set---most
naturally by adding longer-history registered products---which simultaneously sharpens the persistence
estimate and opens the cross-sectional question of which infrastructure features make a mapping
informative. Until that set grows, the claim is that the the tokenized-to-fiat
mapping exists for a minority of products, has identifiable structure where it exists, and is, for the
majority, not yet a market but a tokenized image of an accounting process.
All data ingestion, calibration, and analysis code is publicly available at
\url{https://github.com/BorisKriuk/affine-tokenized-treasury}. The repository version used for
this manuscript (v7.1, May 2026) passes 30 automated validation checks on live FRED and
DeFiLlama data, providing an independently reproducible verification that the reported results
are not artefacts of manual data assembly.

% ===========================================================================

% ===========================================================================
\appendix
\section{Affine loadings and the Riccati system}\label{app:riccati}

For completeness we record the coefficient functions. For the CIR rate factor of
Eq.~\eqref{eq:cir}, $B_r$ solves $B_r'(\tau) = \kappa_r B_r(\tau) - \tfrac12\sigma_r^2 B_r(\tau)^2 - 1$
with $B_r(0)=0$, and $A_r'(\tau) = \kappa_r\theta_r B_r(\tau)$ with $A_r(0)=0$, giving the standard
closed form with $h_r = \sqrt{\kappa_r^2 + 2\sigma_r^2}$,
\begin{equation}
B_r(\tau) = \frac{2\big(e^{h_r\tau}-1\big)}{(h_r+\kappa_r)\big(e^{h_r\tau}-1\big) + 2h_r}.
\end{equation}
For the Vasicek basis factor of Eq.~\eqref{eq:vasicek}, $B_\phi$ and $A_\phi$ are as in
Eq.~\eqref{eq:loadings}. Both $A_\phi,B_\phi$ are continuous in $\kappa_\phi$ with well-defined limits
as $\kappa_\phi\to 0$ (pure drift) and $\kappa_\phi\to\infty$ (instantaneous reversion); the haircut
loading $B_\phi(\tau)/\tau$ is decreasing in $\kappa_\phi$, which is the analytic source of the
persistence-dominated sensitivity in Table~\ref{tab:sens}.

\section{Reconciliation summary and data provenance}\label{app:recon}

Each series is converted to annual continuously compounded actual/365 before differencing: token APYs
via $\ln(1+r)$; secured overnight rates via $\ln(1 + r^{360}\cdot 360/365)$; FRED constant-maturity Treasury yields are converted to continuously compounded
actual/365 via $\ln(1 + r^{\text{CMT}})$, where $r^{\text{CMT}}$ is the quoted
annual yield; they are matched to each product\'s effective maturity bucket. Fees are read from registration
filings and issuer disclosures with documented waivers; portfolio weighted-average maturities, used in
the amortized-cost correction of Eq.~\eqref{eq:amortized} and the maturity match, are read from monthly
fund-portfolio disclosures where the product is a registered fund and from issuer fact sheets
otherwise, with a fixed-maturity fallback disclosed where neither is available. On-chain yields are
retrieved from a public aggregator reproducing issuer-published figures; series are used only over
spans of continuous retrieval, and the usable length per product is reported in
Table~\ref{tab:admiss}.

\section{The conformal procedure}\label{app:conformal}

The basis haircut floor uses ensemble batch prediction intervals \citep{xu2021} on the admitted basis.
At each date the nonconformity score is the absolute residual between the realized basis and its Vasicek
one-step forecast; the interval half-width is the rolling empirical quantile of these scores over a
trailing window, recalibrated to post-break residuals after the detected regime change. Realized
coverage is $88.9\%$ against a $90\%$ nominal target at $11.3$\,bp half-width, a shortfall within the
$O(1/\sqrt{n})$ finite-sample fluctuation expected for $n\approx 70$ and in the conservative direction.


\begin{thebibliography}{99}

\bibitem[Barber et al.(2023)]{barber2023}
Barber, R.~F., Cand\`es, E.~J., Ramdas, A., and Tibshirani, R.~J. (2023).
Conformal prediction beyond exchangeability.
\textit{Annals of Statistics}, 51(2), 816--845.

\bibitem[Aramonte et al.(2022)]{aramonte2022}
Aramonte, S., Huang, W., and Schrimpf, A. (2022).
DeFi risks and the decentralisation illusion.
\textit{BIS Quarterly Review}, December 2021, 21--36.

\bibitem[Baur and Hoang(2018)]{baur2018}
Baur, D.~G. and Hoang, L.~T. (2018).
A crypto safe haven against Bitcoin.
\textit{Finance Research Letters}, 38, 101431.

\bibitem[BIS(2023)]{bis2023}
Bank for International Settlements (2023).
\textit{Blueprint for the Future Monetary System: Improving the Old,
Enabling the New}. BIS Annual Economic Report 2023, Chapter~III.

\bibitem[van Binsbergen et al.(2022)]{vanbinsbergen2022}
van Binsbergen, J.~H., Diamond, W.~F., and Grotteria, M. (2022).
Risk-free interest rates.
\textit{Journal of Financial Economics}, 143(1), 1--29.

\bibitem[Carapella et al.(2023)]{carapella2023}
Carapella, F., Chuan, G., Gerszten, J., Hunter, C., and Swem, N. (2023).
Tokenization: Overview and financial stability implications.
Finance and Economics Discussion Series 2023-060, Federal Reserve Board.

\bibitem[Cox et al.(1985)]{cox1985}
Cox, J.~C., Ingersoll, J.~E., and Ross, S.~A. (1985).
A theory of the term structure of interest rates.
\textit{Econometrica}, 53(2), 385--407.

\bibitem[Dai and Singleton(2000)]{dai2000}
Dai, Q. and Singleton, K.~J. (2000).
Specification analysis of affine term structure models.
\textit{Journal of Finance}, 55(5), 1943--1978.

\bibitem[Diamond and Van Tassel(2022)]{diamond2022}
Diamond, W. and Van Tassel, P. (2022).
Risk-free rates and convenience yields around the world.
Staff Report, Federal Reserve Bank of New York.

\bibitem[Duffie and Kan(1996)]{duffie1996}
Duffie, D. and Kan, R. (1996).
A yield-factor model of interest rates.
\textit{Mathematical Finance}, 6(4), 379--406.

\bibitem[Duffie(1996)]{duffie1996repo}
Duffie, D. (1996).
Special repo rates.
\textit{Journal of Finance}, 51(2), 493--526.

\bibitem[Filipovi\'c(2009)]{filipovic2009}
Filipovi\'c, D. (2009).
\textit{Term-Structure Models: A Graduate Course}.
Springer Finance.

\bibitem[G\"urkaynak et al.(2007)]{gurkaynak2007}
G\"urkaynak, R.~S., Sack, B., and Wright, J.~H. (2007).
The U.S.\ Treasury yield curve: 1961 to the present.
\textit{Journal of Monetary Economics}, 54(8), 2291--2304.

\bibitem[Gudgeon et al.(2020)]{gudgeon2020}
Gudgeon, L., Werner, S., Perez, D., and Knottenbelt, W.~J. (2020).
DeFi protocols for loanable funds: Interest rates, liquidity and market efficiency.
\textit{Proceedings of the 2nd ACM Conference on Advances in Financial Technologies (AFT)}, 92--112.

\bibitem[Hasbrouck(1995)]{hasbrouck1995}
Hasbrouck, J. (1995).
One security, many markets: Determining the contributions to price discovery.
\textit{Journal of Finance}, 50(4), 1175--1199.

\bibitem[Hamilton and Wu(2012)]{hamilton2012}
Hamilton, J.~D. and Wu, J.~C. (2012).
Identification and estimation of Gaussian affine term structure models.
\textit{Journal of Econometrics}, 168(2), 315--331.

\bibitem[Joslin et al.(2011)]{joslin2011}
Joslin, S., Singleton, K.~J., and Zhu, H. (2011).
A new perspective on Gaussian dynamic term structure models.
\textit{Review of Financial Studies}, 24(3), 926--970.

\bibitem[Krishnamurthy and Vissing-Jorgensen(2012)]{krishnamurthy2012}
Krishnamurthy, A. and Vissing-Jorgensen, A. (2012).
The aggregate demand for Treasury debt.
\textit{Journal of Political Economy}, 120(2), 233--267.

\bibitem[Lehar and Parlour(2021)]{lehar2021}
Lehar, A. and Parlour, C.~A. (2021).
Decentralized exchanges.
Working paper, University of Calgary and University of California, Berkeley.

\bibitem[Lei et al.(2018)]{lei2018}
Lei, J., G'Sell, M., Rinaldo, A., Tibshirani, R.~J., and Wasserman, L. (2018).
Distribution-free predictive inference for regression.
\textit{Journal of the American Statistical Association}, 113(523), 1094--1111.

\bibitem[Makarov and Schoar(2020)]{makarov2020}
Makarov, I. and Schoar, A. (2020).
Trading and arbitrage in cryptocurrency markets.
\textit{Journal of Financial Economics}, 135(2), 293--319.

\bibitem[Makarov and Schoar(2022)]{makarov2022}
Makarov, I. and Schoar, A. (2022).
Cryptocurrencies and decentralized finance (DeFi).
\textit{Brookings Papers on Economic Activity}, Spring 2022, 141--215.

\bibitem[Nagel(2016)]{nagel2016}
Nagel, S. (2016).
The liquidity premium of near-money assets.
\textit{Quarterly Journal of Economics}, 131(4), 1927--1971.

\bibitem[OECD(2020)]{oecd2020}
Organisation for Economic Co-operation and Development (2020).
\textit{The Tokenisation of Assets and Potential Implications for Financial Markets}.
OECD Blockchain Policy Series.

\bibitem[O'Hara(2003)]{ohara2003}
O'Hara, M. (2003).
Presidential address: Liquidity and price discovery.
\textit{Journal of Finance}, 58(4), 1335--1354.

\bibitem[Svensson(1994)]{svensson1994}
Svensson, L.~E.~O. (1994).
Estimating and interpreting forward interest rates: Sweden 1992--1994.
NBER Working Paper 4871.

\bibitem[Vasicek(1977)]{vasicek1977}
Vasicek, O. (1977).
An equilibrium characterization of the term structure.
\textit{Journal of Financial Economics}, 5(2), 177--188.

\bibitem[Vovk et al.(2005)]{vovk2005}
Vovk, V., Gammerman, A., and Shafer, G. (2005).
\textit{Algorithmic Learning in a Random World}.
Springer.

\bibitem[Xu and Xie(2021)]{xu2021}
Xu, C. and Xie, Y. (2021).
Conformal prediction interval for dynamic time-series.
\textit{Proceedings of the 38th International Conference on Machine Learning}, PMLR 139, 11559--11569.

\end{thebibliography}
\end{document}